\documentclass{aastex63}
\usepackage{comment}
\usepackage{xcolor}
\newcommand{\RomanNumeralCaps}[1]
    {\MakeUppercase{\romannumeral #1}}

\received{}
\revised{}
\accepted{}
\graphicspath{{./}{figures/}}

\begin{document}

\title{\textit{NuSTAR} Observation of a Minuscule Microflare in a Solar Active Region}

\correspondingauthor{Kristopher Cooper}
\email{k.cooper.2@research.gla.ac.uk}

\author[0000-0001-8589-3378]{Kristopher Cooper}
\affiliation{School of Physics \& Astronomy, University of Glasgow, Glasgow G12 8QQ, UK}

\author[0000-0003-1193-8603]{Iain G. Hannah}
\affiliation{School of Physics \& Astronomy, University of Glasgow, Glasgow G12 8QQ, UK}

\author[0000-0002-1984-2932]{Brian W. Grefenstette}
\affiliation{Cahill Center for Astrophysics, 1216 E. California Blvd, California Institute of Technology, Pasadena, CA 91125, USA}

\author[0000-0001-7092-2703]{Lindsay Glesener}
\affiliation{School of Physics \& Astronomy, University of Minnesota—Twin Cities, Minneapolis, MN 55455, USA}

\author{S\"{a}m Krucker}
\affiliation{University of Applied Sciences and Arts Northwestern Switzerland, 5210 Windisch, Switzerland}
\affiliation{Space Sciences Laboratory University of California, Berkeley, CA 94720, USA}

\author[0000-0001-5685-1283]{Hugh S. Hudson}
\affiliation{School of Physics \& Astronomy, University of Glasgow, Glasgow G12 8QQ, UK}
\affiliation{Space Sciences Laboratory University of California, Berkeley, CA 94720, USA}

\author[0000-0002-8574-8629]{Stephen M. White}
\affiliation{Air Force Research Laboratory, Space Vehicles Directorate, Kirtland AFB, NM 87123, USA}

\author{David M. Smith}
\affiliation{Santa Cruz Institute of Particle Physics and Department of Physics, University of California, Santa Cruz, CA 95064, USA}



\begin{abstract}

We present X-ray imaging spectroscopy of one of the weakest active region (AR) microflares ever studied. The microflare occurred at $\sim$11:04~UT on 2018 September 9 and we studied it using the \textit{Nuclear Spectroscopic Telescope ARray} (\textit{NuSTAR}) and the \textit{Solar Dynamic Observatory's} Atmospheric Imaging Assembly (\textit{SDO}/AIA). The microflare is observed clearly in 2.5--7~keV with \textit{NuSTAR} and in Fe~\RomanNumeralCaps{18} emission derived from the hotter component of the 94~\AA{} \textit{SDO}/AIA channel. We estimate the event to be three orders of magnitude lower than a \textit{GOES} A class microflare with an energy of 1.1$\times$10$^{26}$~erg. It reaches temperatures of 6.7~MK with an emission measure of 8.0$\times$10$^{43}$~cm$^{-3}$. Non-thermal emission is not detected but we instead determine upper limits to such emission. We present the lowest thermal energy estimate for an AR microflare in literature, which is at the lower limits of what is still considered an X-ray microflare.

\end{abstract}

\keywords{Sun: activity --- Sun: corona --- Sun: flares --- Sun: X-rays, gamma rays}






\section{Introduction} \label{sec:intro}
Solar flares occur in active regions (ARs) and rapidly release stored magnetic energy into heating, mass flows, and particle acceleration in its vicinity \citep{benz_flare_2017}. The energy released can vary greatly, with smaller solar flares (called microflares) having energies about 10$^{28}$--10$^{26}$~erg \citep{lin_solar_1984, hannah_microflares_2011}. Microflares have been extensively studied in X-rays, to determine their thermal and non-thermal properties and are observed to have GOES soft X-ray fluxes $<$10$^{-6}$~W~m$^{-2}$ and so are B, A, or sub-A class flares. Even smaller events (called nanoflares) were proposed by \cite{parker_nanoflares_1988} as a unit of impulsive energy release to heat the whole corona, not just in ARs. These incredibly small events, with energies about 10$^{24}$~erg, would need to be highly frequent with their frequency distribution having a power-law index $>$2 to dominate energetically over the larger flares \citep{crosby_frequency_1993, hudson_solar_1991}. The term nanoflare is sometimes used to describe an observed extreme-ultraviolet (EUV) brightening with energies about this scale.

X-ray emission from microflares has been extensively studied in the past with instruments such as the \textit{Reuven Ramaty High-Energy Solar Spectroscopic Imager} (\textit{RHESSI}, \cite{lin_reuven_2002}). A comprehensive study of more than 25,000 microflaring events observed by \textit{RHESSI} found that they shared many properties with their larger counterparts \citep{christe_rhessi_2008, hannah_rhessi_2008}. It was also noted that physical flare size did not seem to correlate with the magnitude of the microflare. To extend this work to even smaller flares requires improved sensitivity.

The \textit{Nuclear Spectroscopic Telescope ARray} (\textit{NuSTAR}) is an X-ray astrophysical telescope with the capability of observing the Sun above 2.5~keV with unprecedented sensitivity \citep{harrison_nuclear_2013}. \textit{NuSTAR} consists of Wolter-\RomanNumeralCaps{1} type optics on a 10~m mast that focus X-rays onto two focal plane modules (FPMA and FPMB), each with a field of view of 12$'\times$12$'$, made up of four pixelated CdZnTe detectors separated by chip gaps. \textit{NuSTAR} detects individual counts, with a throughput of 400~counts~s$^{-1}$~module$^{-1}$. Even quiet or weakly flaring emission from the Sun can produce high count rates, resulting in significant deadtime and low effective exposure, thus most solar observations operate with a livetime fraction $\ll$1 \citep{grefenstette_first_2016}. This can limit \textit{NuSTAR}'s sensitivity to the hottest material or weaker non-thermal energy during periods when livetime is small.

Since the first solar NuSTAR observations in 2014 \citep[see][]{grefenstette_first_2016, hannah_first_2016, kuhar_evidence_2017}, solar activity has decreased allowing sub-A class microflares to be observed regularly within ARs \citep{glesener_nustar_2017, glesener_accelerated_2020, wright_microflare_2017, hannah_joint_2019} and small brightenings outside of ARs \citep{kuhar_nustar_2018}\footnote{Overview of all \textit{NuSTAR} solar observations available at \url{https://ianan.github.io/nsigh_all/}}. The AR microflares observed by \textit{NuSTAR} have been found to have thermal energy releases down to 10$^{27}$~erg with quiet Sun brightenings having energies down to 10$^{26}$~erg. Non-thermal emission has rarely been observable in \textit{NuSTAR} microflare analyses, with \cite{glesener_accelerated_2020} reporting the first focusing optics imaging spectroscopy of non-thermal emission from an A5.7 class microflare. Limits on the non-thermal emission have been determined in other \textit{NuSTAR} microflare analyses \citep{wright_microflare_2017}.

In this Letter, we present observations of a microflare from 2018 September 9 at $\sim$11:04~UT (SOL2018-09-09T11) in AR AR12721. This event was observed with \textit{NuSTAR} and also in EUV with the \textit{Solar Dynamic Observatory}'s Atmospheric Imaging Assembly \citep[\textit{SDO}/AIA;][]{lemen_atmospheric_2012}. In Section \ref{sec:ar_micro} the whole \textit{NuSTAR} campaign, across 2018 September 9--10, is briefly discussed. We then focus on the small microflare's time profiles and spatial properties (Section \ref{sec:lc&im}) followed by X-ray spectral analysis and \textit{GOES} flare classification (Section \ref{sec:spec. fit}). A thermal energy estimate is then calculated and compared to previously obtained values for other microflares in Section \ref{sec:energy}. An upper limit on the non-thermal emission of the microflare is also derived in Section \ref{sec:nonthermal}. In Section \ref{sec:multi-thermal}, \textit{NuSTAR} and \textit{SDO}/AIA loci curves and emission measure distributions are calculated along with a comparison of the emission detected from both observatories.

\section{Weakest AR X-Ray Microflare} \label{sec:ar_micro}
\textit{NuSTAR} performed six hour-long solar observations on 2018 September 9--10 with AR12721 (that emerged September 8) dominating the field of view.. This campaign was related to a region targeted by the \textit{FOXSI-3} sounding rocket \citep{musset_ghost-ray_2019} on September 7, which was still in the \textit{NuSTAR} field of view, but fainter and less active than AR12721. Numerous X-ray microflares produced by AR12721 were seen over the two-day observing window. In this Letter we focus on one of the smaller events; the other microflares are the subjects of a later paper.

\subsection{Time Profile and Imaging} \label{sec:lc&im}
\begin{figure*}
\gridline{\fig{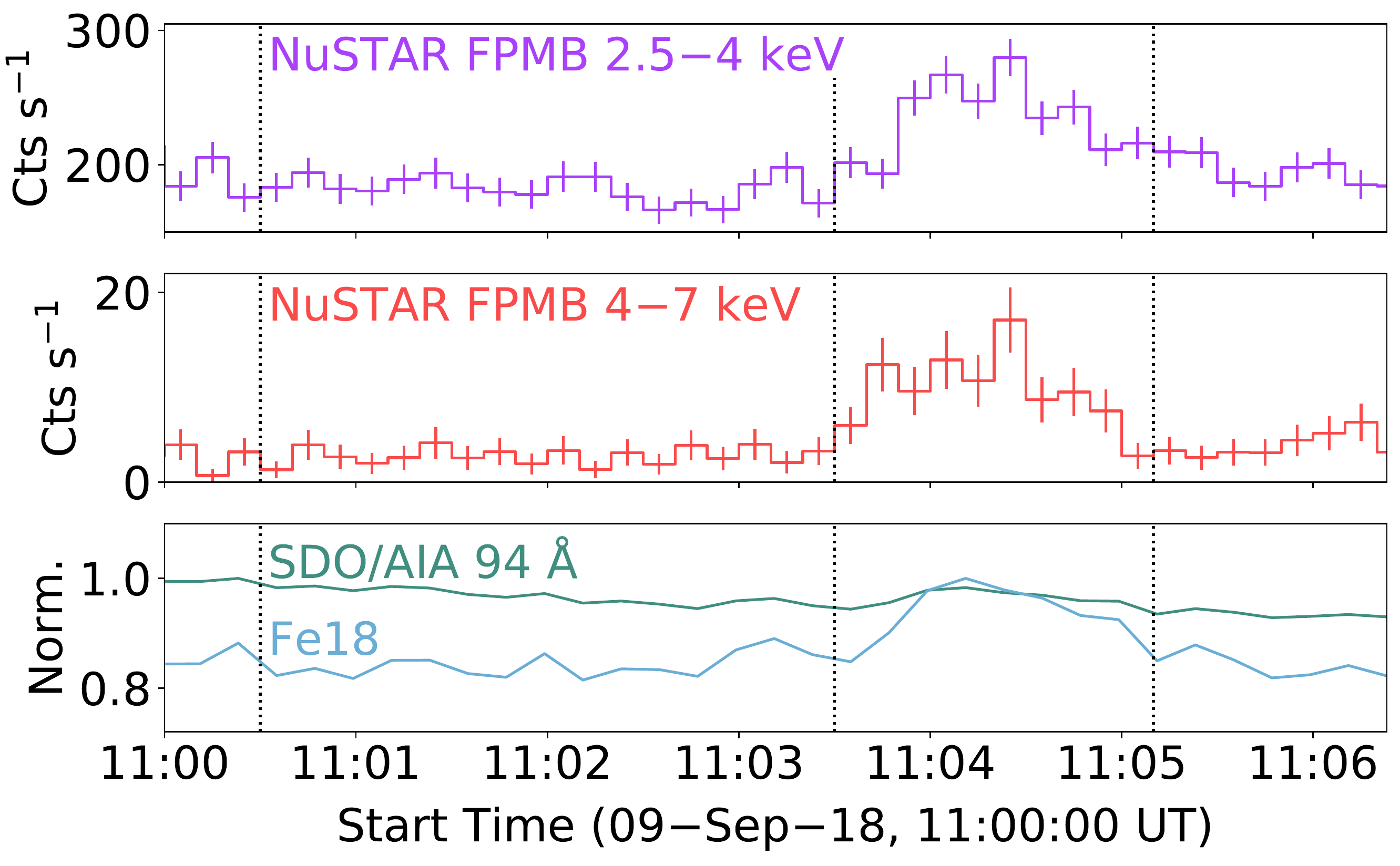}{0.56\textwidth}{(a)}
\fig{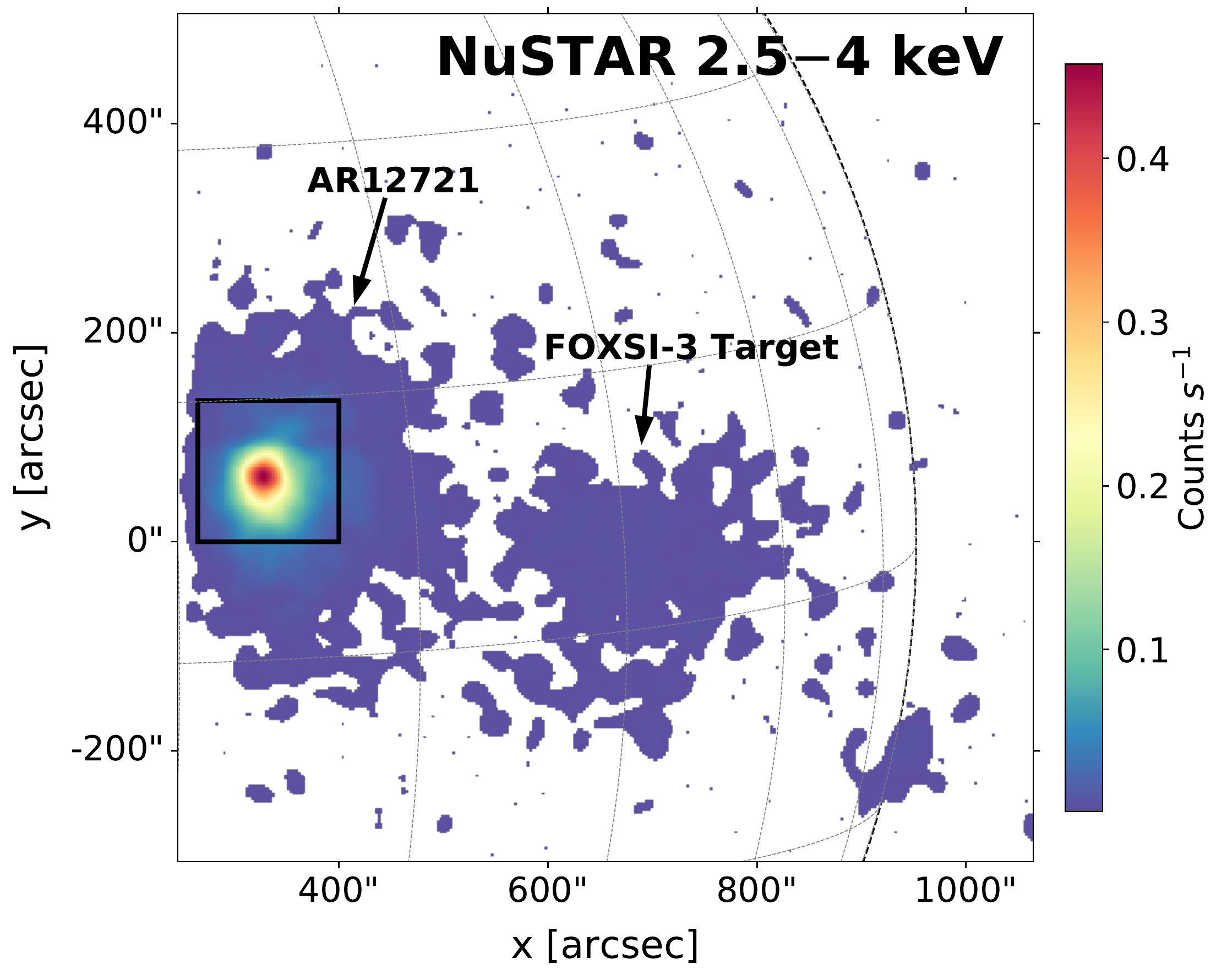}{0.42\textwidth}{(b)}}
\gridline{\fig{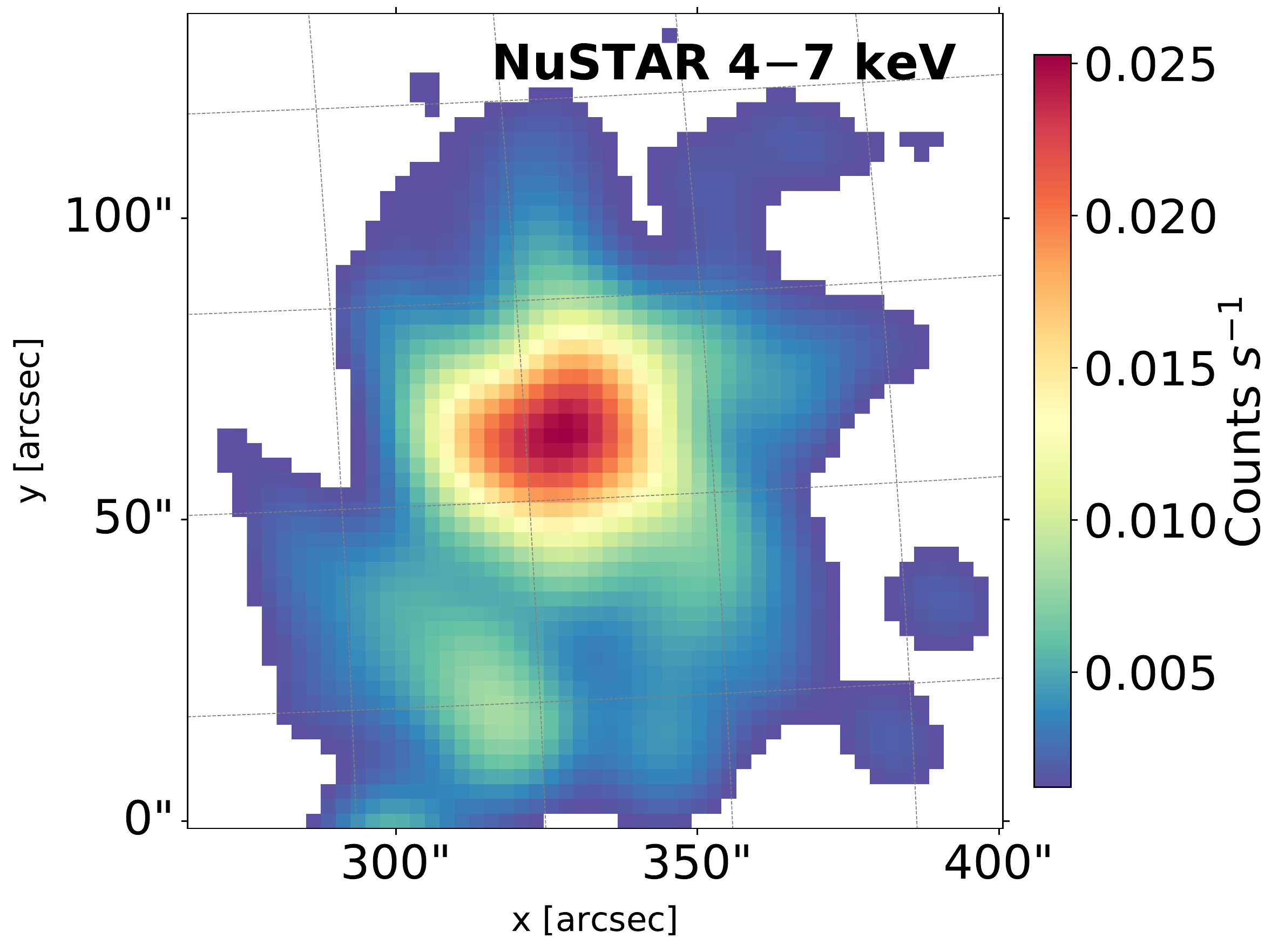}{0.37\textwidth}{(c)}
\fig{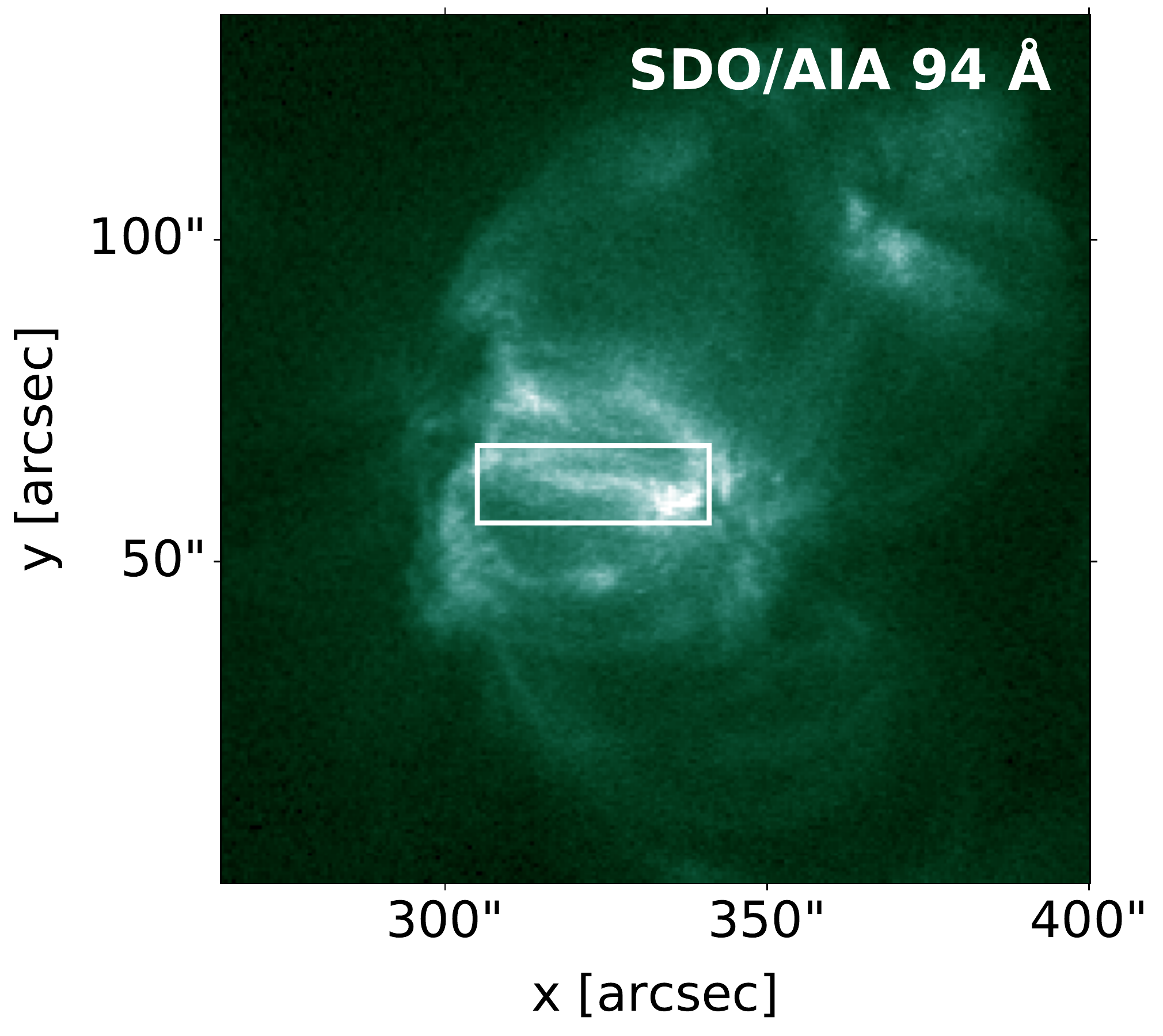}{0.31\textwidth}{(d)}
\fig{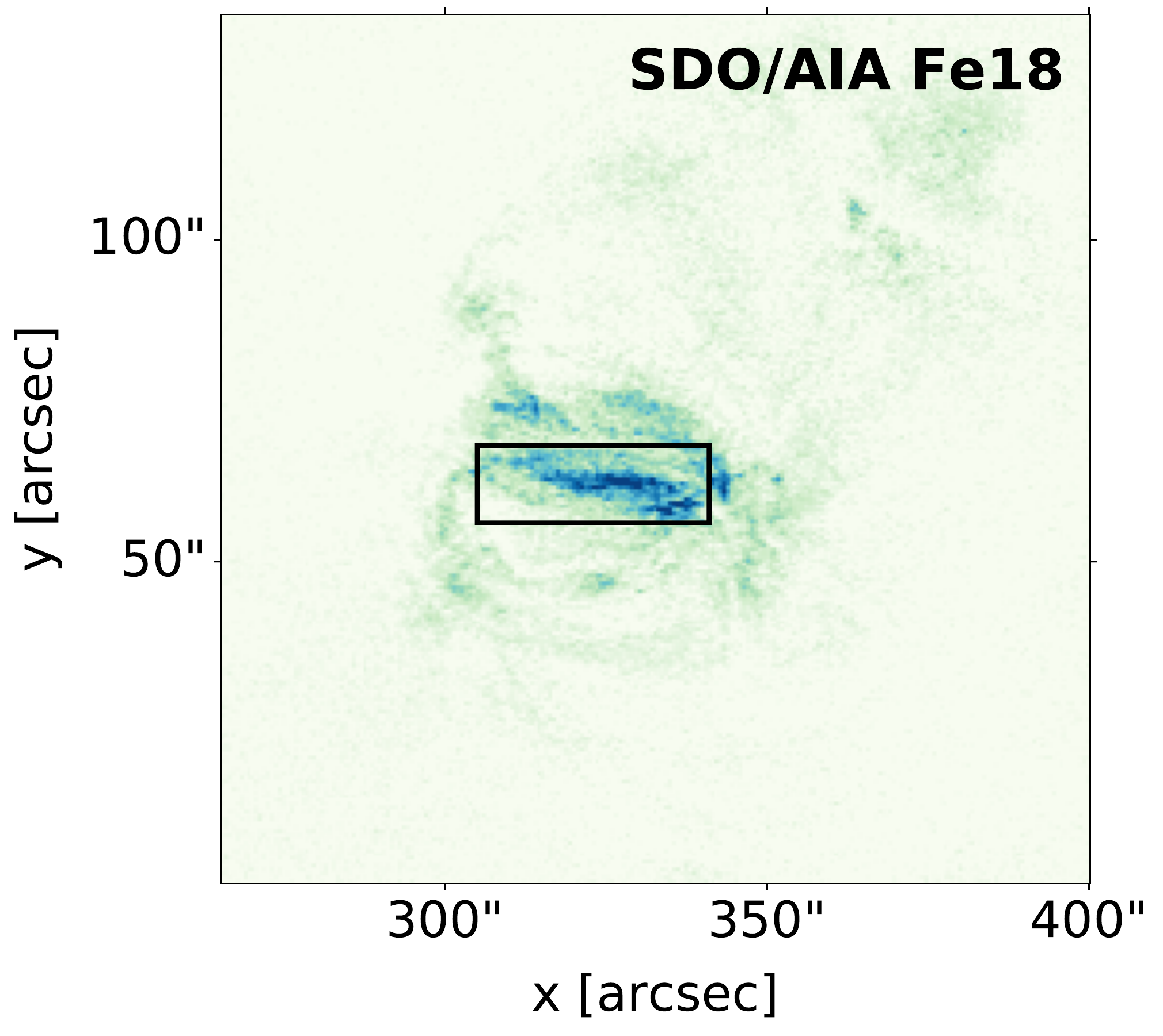}{0.31\textwidth}{(e)}}

\caption{\textit{NuSTAR} 2.5--4~keV and 4--7~keV time profiles with \textit{SDO}/AIA 94~\AA{} and Fe~\RomanNumeralCaps{18} maximum normalized lightcurves of the microflaring event on 2018 September 9 (a). The black box in panel (b) indicates the region used to produce the \textit{NuSTAR} lightcurves (purple, red) and the region shown in panels (c), (d), and (e). The area used to produce the \textit{SDO}/AIA time profiles (green, blue) are indicated in their panels, (d) and (e), with a white and black box, respectively. The vertical dotted lines in panel (a) encase the pre-flare time (11:00:30 to 11:03:30~UT) and the microflaring time (11:03:30 to 11:05:10~UT). \textit{SDO}/AIA and \textit{NuSTAR} images were integrated over the microflaring time. A Gaussian filter with a FWHM of $\sim$15 arcseconds was used to smooth the \textit{NuSTAR} X-ray images that have been livetime corrected.
\label{fig:lightcurves}}
\end{figure*}

The microflare presented was initially discovered upon inspection of the \textit{NuSTAR} lightcurve, shown in Figure~\ref{fig:lightcurves}, panel (a), calculated from the region shown in panel (b). The emission from the microflare becomes more pronounced above the background in the higher energy range of 4--7~keV compared to 2.5--4~keV. A corresponding ``bump" can be seen clearly in the \textit{SDO}/AIA Fe~\RomanNumeralCaps{18} proxy \citep{del_zanna_multi-thermal_2013} but \textit{SDO}/AIA 94~\AA{}, nor the other \textit{SDO}/AIA channels, displayed a clear feature. The 94~\AA{} images show a loop better than any other original SDO/AIA channel, but it is only in the Fe~\RomanNumeralCaps{18} component that there is clear evidence of the microflare heating (Figure~\ref{fig:lightcurves}, panels (d) and (e)). Due to the position of the event on the \textit{NuSTAR} focal plane, only the data obtained from FPMB can be used as the detector chip gap affects the FPMA data. It does, however, provide corroboratory evidence for the event as it also shows a clear microflare time profile.

It should be noted that, as expected of flaring behavior, it appears that the emission seen from the higher NuSTAR energy range, 4--7~keV, rises slightly before emission peaks in lower energies from NuSTAR 2.5--4~keV and Fe~\RomanNumeralCaps{18}. This could be due to hotter plasma earlier in the event or an initially accelerated electron distribution, a potential indication of non-thermal emission. 

The \textit{NuSTAR} pointing only requires a single correction over the entire time of the flare, found by aligning the \textit{NuSTAR} image to \textit{SDO}/AIA. We co-align the \textit{NuSTAR} images with the Fe~\RomanNumeralCaps{18} microflare emission map shown in panel (e). The shift in the \textit{NuSTAR} pointing was obtained by cross-correlating the X-ray and EUV images. Even after the spatial co-alignment, there still remains a conservative shift uncertainty of approximately 10~arcseconds. This is due to the lack of defined structure in the X-ray image. 

\begin{figure*}
\gridline{\fig{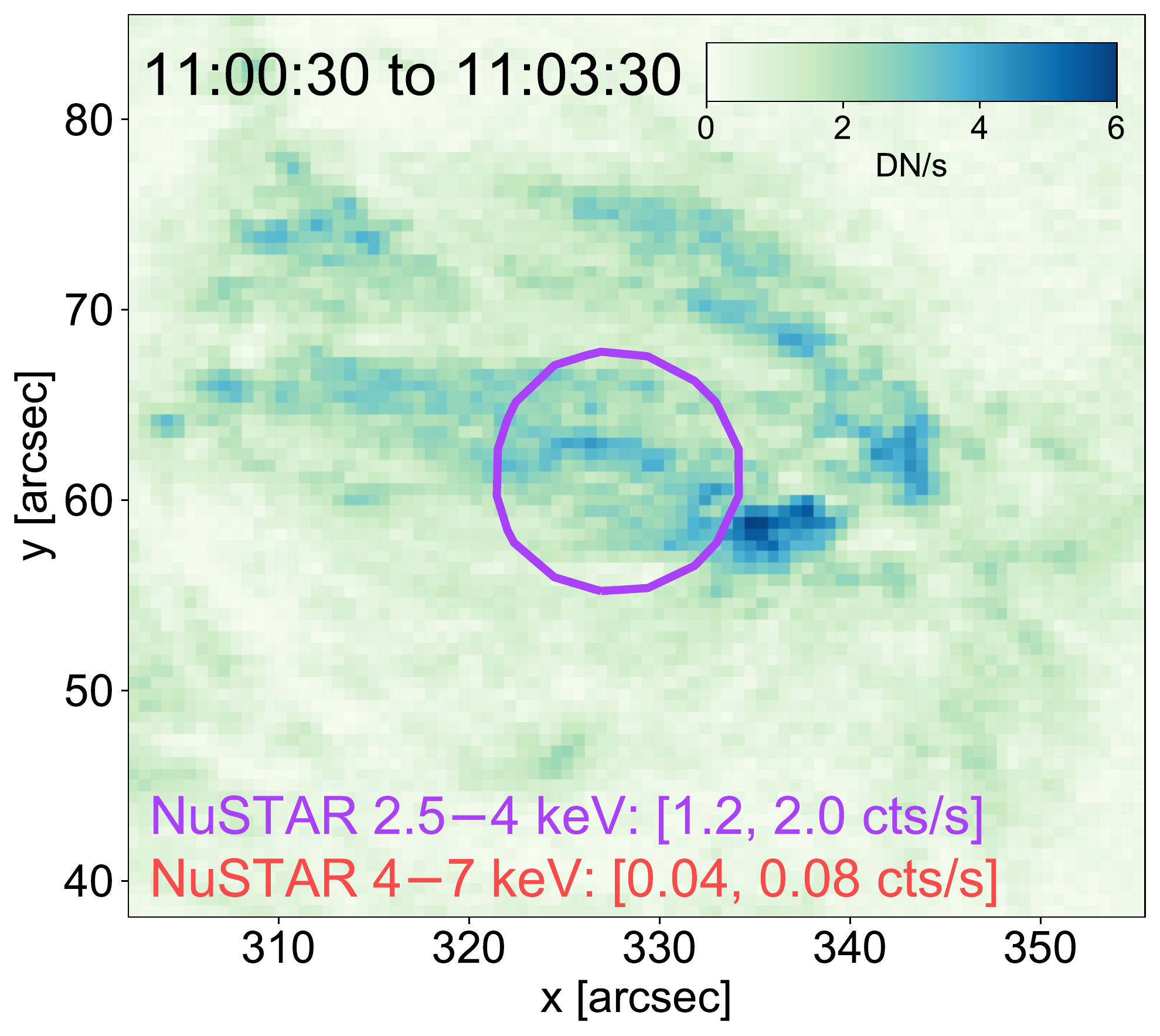}{0.42\textwidth}{}
        \hspace{-1cm}
          \fig{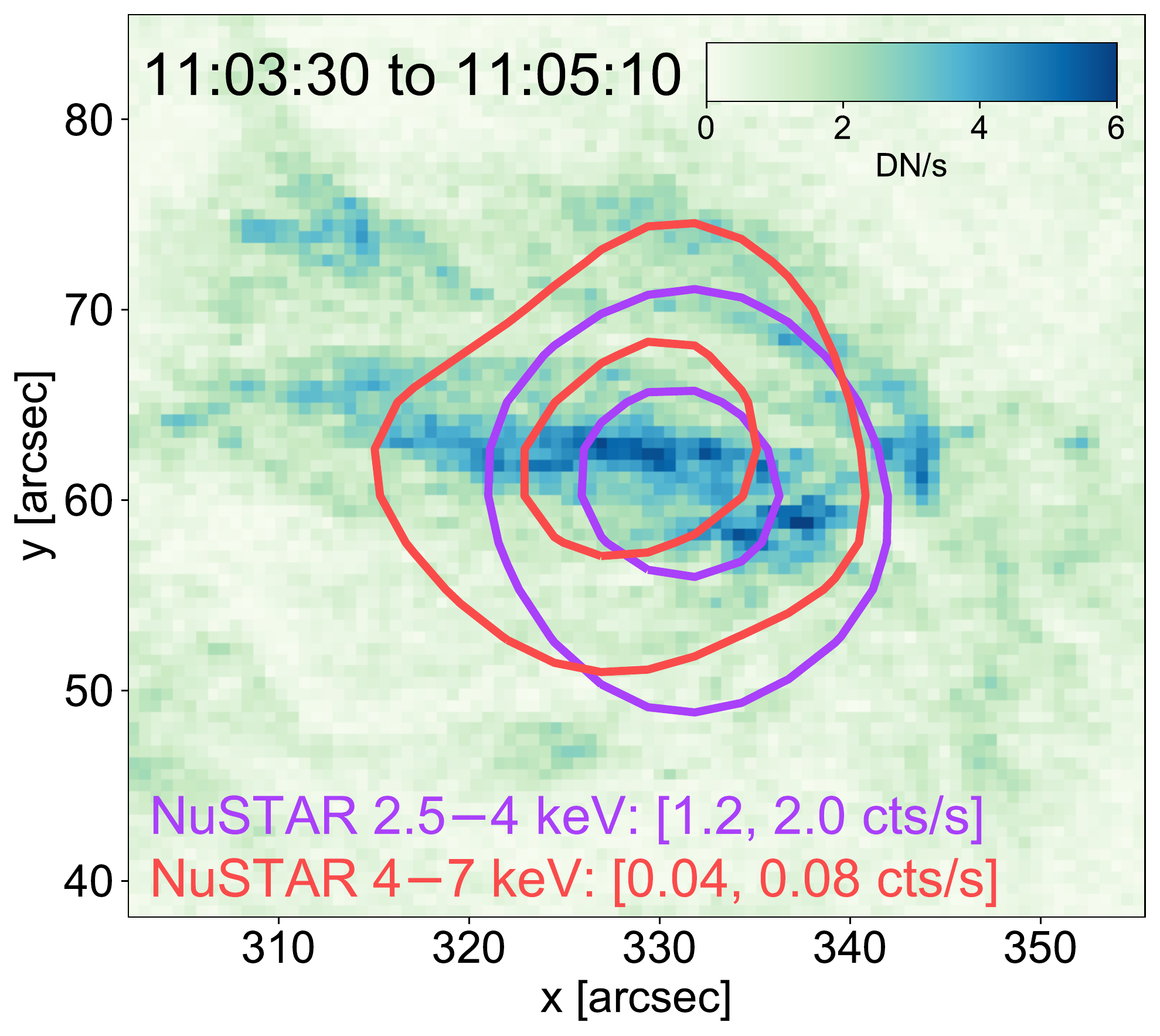}{0.42\textwidth}{}
          }
    \vspace{-0.7cm}
\gridline{\fig{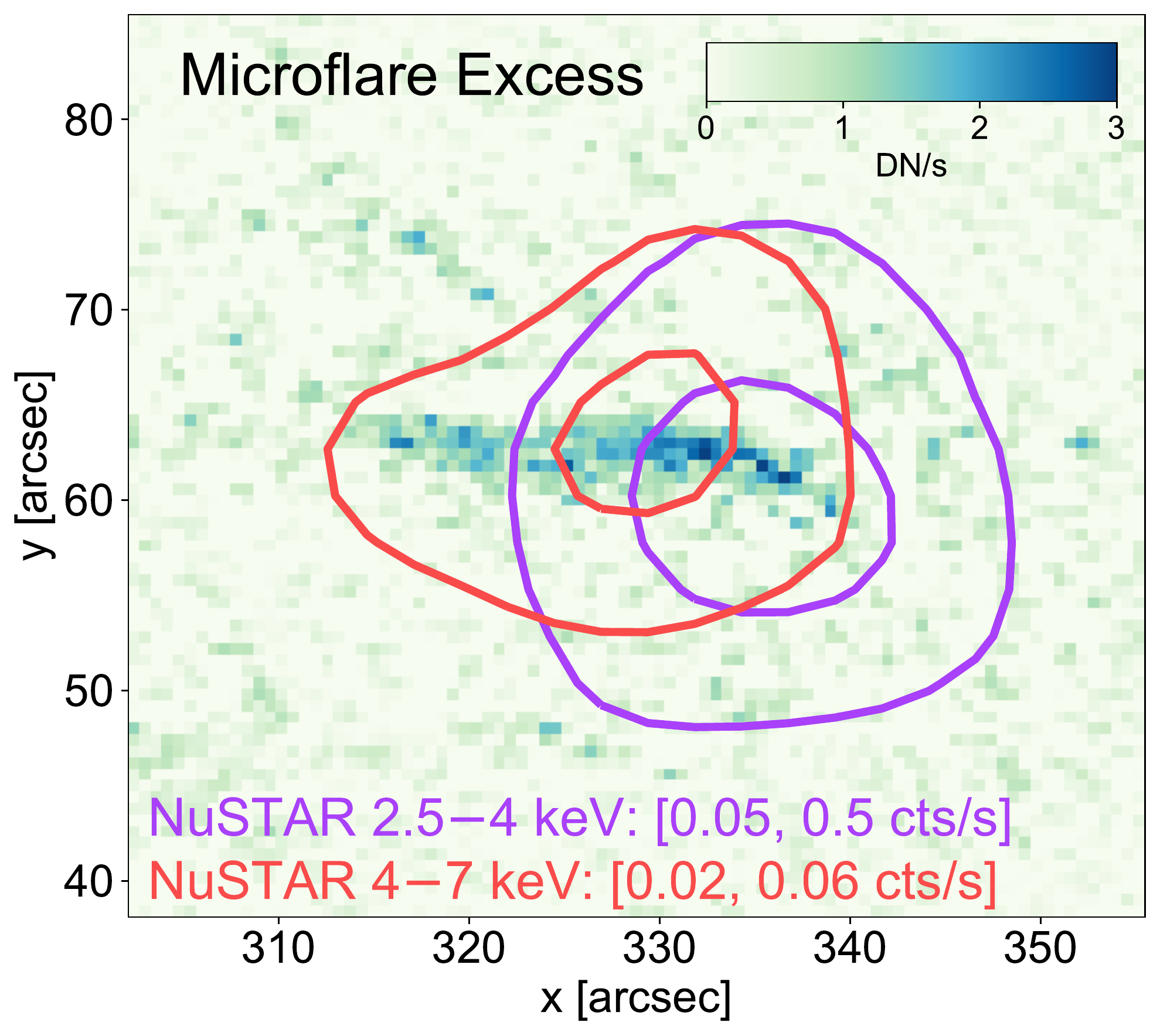}{0.42\textwidth}{}}
    \vspace{-0.8cm}
\caption{\textit{SDO}/AIA Fe~\RomanNumeralCaps{18} map with 2.5--4~keV and 4--7~keV \textit{NuSTAR} absolute contour levels for the pre-flare time (top-left panel) and the microflaring time (top-right panel). The bottom panel shows the the pre-flare subtracted map, i.e. the microflare excess.
\label{fig:contours}}
\end{figure*}

Contour maps of the shifted NuSTAR data on an Fe~\RomanNumeralCaps{18} background during the pre-flare and microflare times are presented in Figure~\ref{fig:contours}. The energy ranges are the same as those used in the lightcurves from Figure~\ref{fig:lightcurves}. Contours created from 2.5--4~keV (purple) and 4--7~keV (red) emission were deconvolved using a Lucy-Richardson method over 20 and 10 iterations, respectively \citep{richardson_bayesian-based_1972}. There is some 2.5--4~keV emission during the pre-flare time, which becomes brighter during the microflare and joined by the 4--7~keV source at the same location. 

To see the heating due to the microflare we subtract the pre-flare image from the microflare one - the resulting microflare excess is shown in Figure \ref{fig:contours} (bottom panel). Here an elongated loop is more clearly visible in Fe~\RomanNumeralCaps{18} and the 2.5--4~keV source is similar to before. However, now the 4--7~keV source is more elongated and the centroid is slightly shifted to the left of the 2.5--4~keV source. This may not be a significant shift as it is within the spatial resolution of \textit{NuSTAR} \citep{grefenstette_first_2016}. Although the time profile (Figure~\ref{fig:lightcurves}) and later the spectral fit results (Figure~\ref{fig:spec_fits}) show a definite but small event, the physical size of the microflaring loop ($\sim$20 arcseconds in length) is not uncommon from others observed in X-rays \citep{glesener_nustar_2017, hannah_rhessi_2008}.

\subsection{NuSTAR Spectral Fitting} \label{sec:spec. fit}
\begin{figure*}
\gridline{\fig{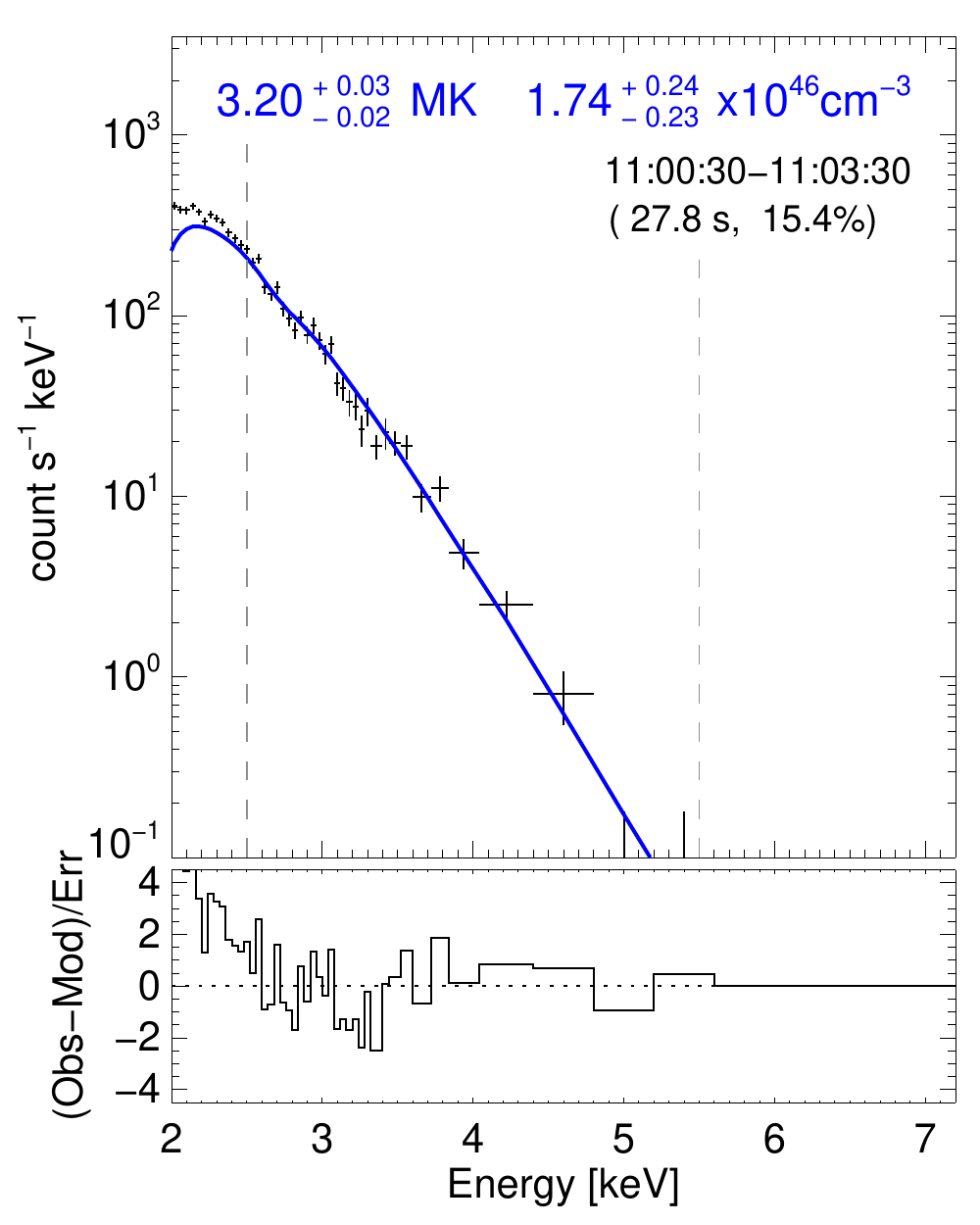}{0.42\textwidth}{}
          \fig{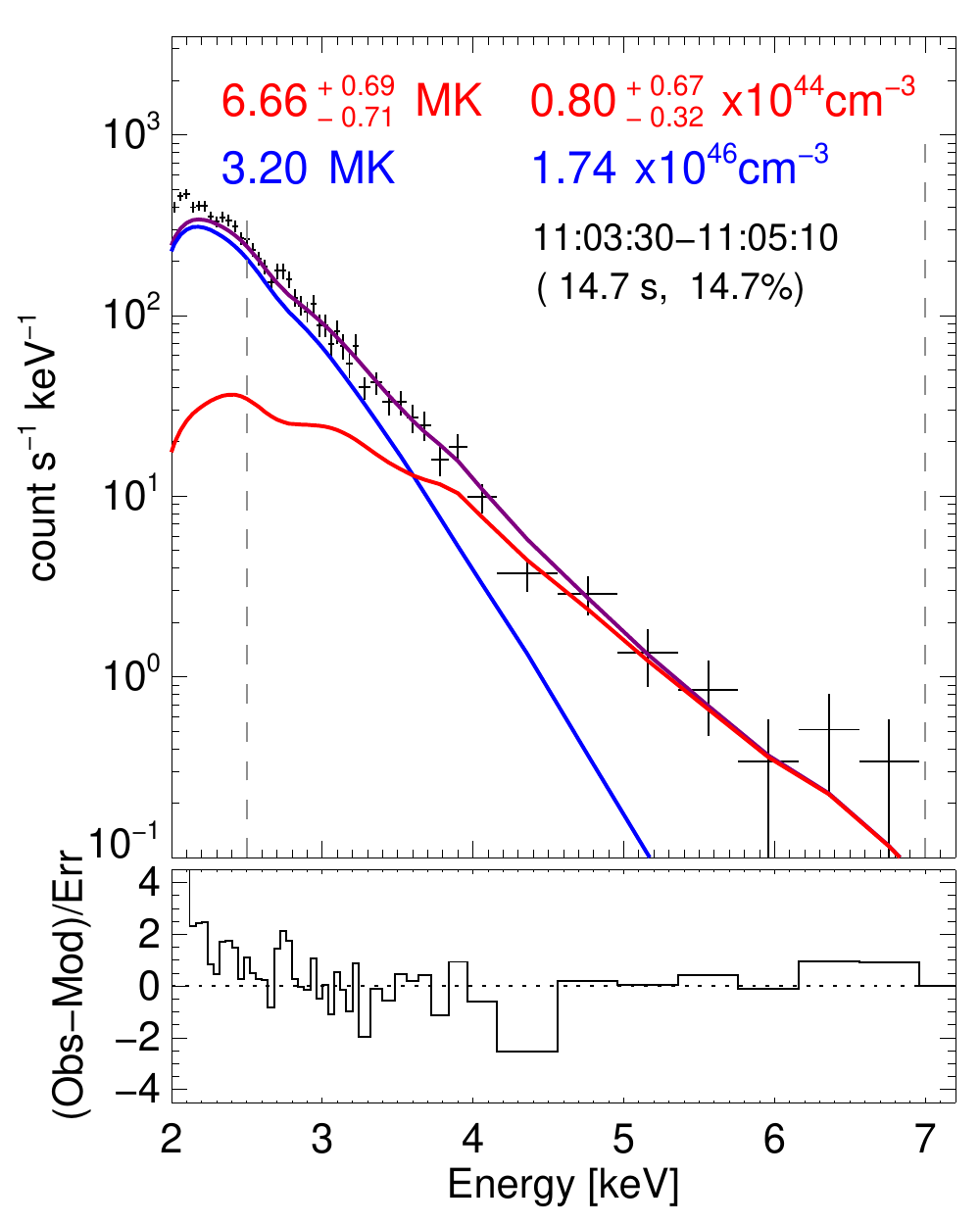}{0.42\textwidth}{}
          }
    \vspace{-0.8cm}
\caption{Thermal model fits, using XSPEC, of \textit{NuSTAR} emission during the pre-flare time (left panel) and microflare time (right panel). The pre-flare spectrum is fitted with one thermal model (blue), which is then used as a fixed component for the microflare spectrum fit along with an additional thermal model (red). Both models in the microflare spectrum combine to give the overall model (purple). The temperatures, emission measures, and times ranges are shown for the spectra with their effective exposures and livetimes in brackets. The quoted errors denote a 90\% confidence range with the fit over the energy range indicate by the vertical dashed lines.
\label{fig:spec_fits}}
\end{figure*}

In order to quantify thermal emission of the AR and microflare we fit the \textit{NuSTAR} FPMB spectra (Figure~\ref{fig:spec_fits}). A circular region, centered on the brightest emission over the pre-flare and microflare times, with a radius of 26.5~arcseconds, was used to produce both spectra.

The pre-flare spectrum (emission from 11:00:30 to 11:03:30~UT, Figure~\ref{fig:spec_fits}, left) is well fitted with a single APEC thermal model using XSPEC \citep{arnaud_astronomical_1996}. The fit gives a temperature of 3.2~MK and an emission measure of 1.7$\times10^{46}$~cm$^{-3}$. These are typical values of quiescent/non-flaring ARs measured by \textit{NuSTAR} \citep{wright_microflare_2017, glesener_nustar_2017, hannah_joint_2019}. However, the livetime recorded throughout this event is significantly larger than those previously studied ($\sim$15\% compared to 1--5\%) resulting in a better sensitivity, and hence constraint, on hotter material. This spectral fit was used as a fixed component during the microflare time (11:03:30 to 11:05:10~UT, Figure~\ref{fig:spec_fits}, right panel) with an additional APEC thermal component used to fit the microflare excess. 

The microflare excess has a harder spectrum that dominates over the pre-flare emission $>$4~keV, similar to what was found in Section \ref{sec:lc&im}. The event excess was fitted with a temperature of 6.7~MK and an emission measure of 8.0$\times10^{43}$~cm$^{-3}$. The temperature is similar, or slightly hotter, to those found from other weak microflaring events, whereas the emission measure is an order of magnitude smaller \citep{glesener_nustar_2017, hannah_joint_2019}. 

The excess thermal fit does not change considerably when taking into account the uncertainty in the pre-flare model. It should be noted that because temperature and emission measure are inversely correlated, the highest/lowest temperature corresponds to the lowest/highest emission measure with asymmetric errors on each. We find that the temperature derived for the microflare excess through spectral fitting is consistent with the presence of emission in the \textit{SDO}/AIA Fe~\RomanNumeralCaps{18} channel. 

Using the \verb|goes_flux49.pro|\footnote{\url{https://hesperia.gsfc.nasa.gov/ssw/gen/idl/synoptic/goes/goes_flux49.pro}} IDL routine in conjunction with the temperature and emission measure of the microflare excess, it is possible to obtain an estimated \textit{GOES} classification for the event. We find a flux of 5$\times10^{-11}$~W~m$^{-2}$, an equivalent \textit{GOES} class of $\sim$A0.005.

\subsection{Thermal Energy}  \label{sec:energy}
From the temperature ($T$) and emission measure ($EM$) of the microflare excess, with the addition of a volume estimate ($V$) for the heated loop, the instantaneous thermal energy ($E_{th}$) can be calculated as

\begin{equation} \label{eq:thermal_energy}
    E_{th} = 3 N_{e} k_{B} T = 3(V \times EM)^{\frac{1}{2}} k_{B} T \hspace{0.5cm}\text{[erg]},
\end{equation}

\noindent
where $N_{e}$ is the total number of thermal electrons and $k_{B}$ is Boltzmann's constant \citep{hannah_rhessi_2008}. The temperature and emission measure are taken from the microflare excess, given in Figure~\ref{fig:spec_fits} (right panel). The volume of the loop can be estimated from the EUV \textit{SDO}/AIA Fe~\RomanNumeralCaps{18} image (Figure~\ref{fig:contours}, bottom panel).

The microflaring loop appears to be 22~by~2~arcseconds (approximately 1.6$\times10^{9}$~by~1.3$\times10^{8}$~cm). Taking the heated loop as a half-torus, this gives a volume of 3.2$\times10^{25}$~cm$^{3}$. Thus, using Equation \ref{eq:thermal_energy} with a temperature of 6.7~MK and emission measure 8.0$\times10^{43}$~cm$^{-3}$, we find a thermal energy of 1.4$^{+0.3}_{-0.2}\times$10$^{26}$~erg. The volume estimated here is undoubtedly an upper limit as it could be contested that the region in EUV is smaller still as most of the emission appears to be focused at the right of the loop with surrounding fainter emission. In addition, this volume estimate does not consider a loop-filling factor, making the thermal energy estimate an upper limit. This thermal energy value is lower than the previous smallest observed \textit{NuSTAR} microflare \citep{hannah_joint_2019}, which was cooler but with a higher emission measure and had a \textit{GOES} class of A0.02. EUV observations of magnetically braided loops observed heating with thermal energy of about 10$^{26}$~erg \citep{cirtain_energy_2013}; however, this was for material up to 4~MK.

\subsection{Non-thermal Limits}  \label{sec:nonthermal}
Following the approach in \cite{wright_microflare_2017}, it is possible to obtain upper limits on any non-thermal emission produced by the microflare from \textit{NuSTAR}'s spectral response. This is done by adding a thick target model (\verb|f_thick2.pro|\footnote{\url{https://hesperia.gsfc.nasa.gov/ssw/packages/xray/idl/f_thick2.pro}}) to a simulated spectrum obtained from the total microflare thermal model. This non-thermal model depends on the power-law index, the low-energy cut-off, and the electron flux of the microflare accelerated electrons. The non-thermal power was calculated throughout this parameter space, where the thick target model gave fewer than four counts at energies greater than 7~keV---consistent with a null detection to 2$\sigma$ \citep{gehrels_confidence_1986}---and that the introduced non-thermal counts were within Poissonian uncertainty at energies $\leqslant$7~keV. We find that the upper non-thermal limits produced are consistent with the required heating over the microflare time ($\sim$10$^{24}$~erg~s$^{-1}$) but only with low-energy cut-offs down to $\sim$6~keV with a power-law index $\geqslant$6. 

The upper limit values calculated that satisfy this microflare heating are lower than the upper limits in \cite{wright_microflare_2017}. This is expected as the microflare discussed here is less energetic. The largest upper limit obtained from this analysis ($\sim$10$^{25}$~erg~s$^{-1}$) is only just comparable to the smallest non-thermal power in similar sized microflares \citep[Table 1]{hannah_rhessi_2008}. The power-law index and cut-off energy are consistent with the values obtained in \cite{glesener_accelerated_2020}. Only the electron flux is different ($\sim$10$^{3}$~larger) which could be expected as the peak emission is also orders of magnitude larger than the flare discussed here. However, the values obtained are not consistent with the events presented in \cite{testa_evidence_2014} who investigated coronal loop footpoint brightenings in iltraviolet (UV) and a nanoflare heating model. Their model required that an electron distribution with a higher low-energy cut-off ($\sim$10 keV) to match their observations compared to the microflare presented.

\subsection{Multi-thermal Microflare Analysis} \label{sec:multi-thermal}
\begin{figure*}
\gridline{\fig{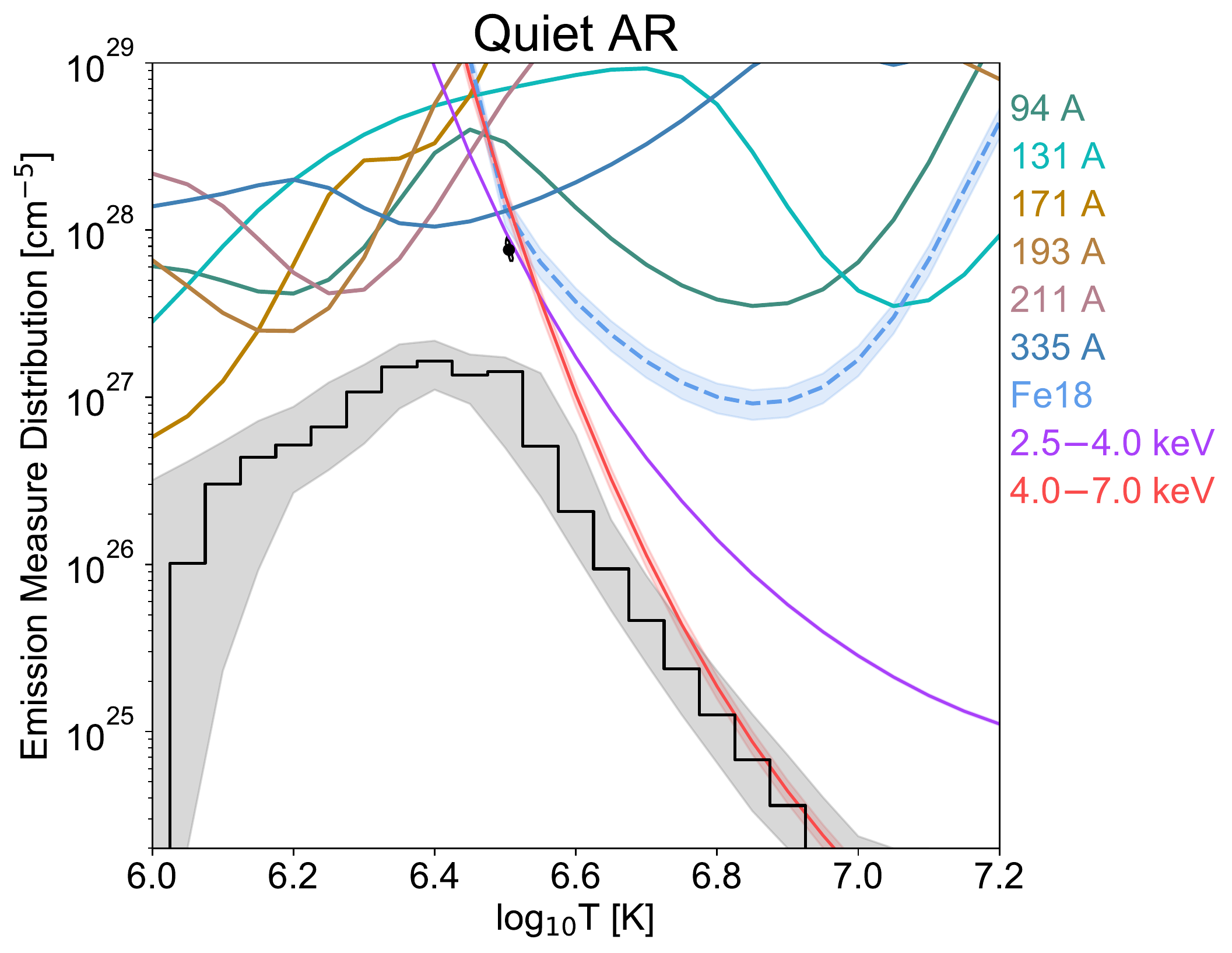}{0.48\textwidth}{}
        \hspace{-1cm}
          \fig{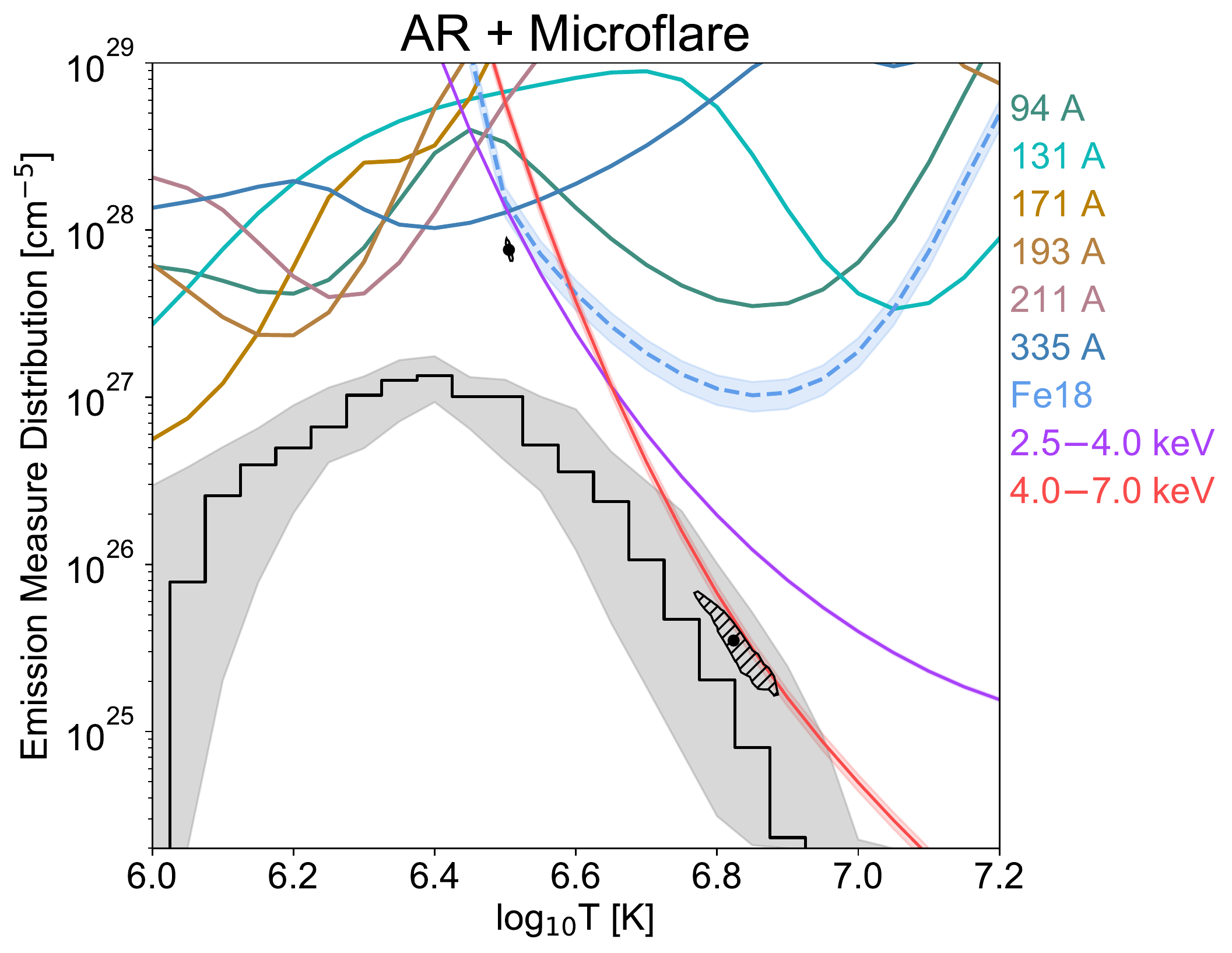}{0.48\textwidth}{}
          }
    \vspace{-0.8cm}
\caption{\textit{NuSTAR} and \textit{SDO}/AIA loci curve plots with the calculated emission measure distributions (black) for the pre-flare time (left panel) and microflare time (right panel). The shaded areas denote the uncertainty range for the \textit{NuSTAR} loci curves (purple and red), the Fe~\RomanNumeralCaps{18} curve (dashed, blue), and the emission measure distribution (gray). The region used to calculate the \textit{SDO}/AIA and \textit{NuSTAR} instrument loci curves was the boxed region shown in Figure~\ref{fig:lightcurves}, panels (d) and (e). The spectral fit values for both times are indicated with their hatched 90\% confidence regions.
\label{fig:emd}}
\end{figure*}

Figure~\ref{fig:emd} shows the EM loci curves (flux divided by temperature response) from \textit{SDO}/AIA and \textit{NuSTAR} plotted with the temperatures and emission measures obtained from Figure~\ref{fig:spec_fits} and their 90\% confidence region (hatched regions). During the pre-flare time (Figure~\ref{fig:emd}, left), the Fe~\RomanNumeralCaps{18} and \textit{NuSTAR} loci curves almost intersect at the temperature and emission measure from the spectral fit. This indicates that similar emission is observed by \textit{NuSTAR} and Fe~\RomanNumeralCaps{18} at the pre-flare stage over the selected region \citep{hannah_joint_2019}. The microflare time has the additional heated plasma from the flaring process (see Figure~\ref{fig:spec_fits} and Figure~\ref{fig:emd}, right) which as expected, results in Fe~\RomanNumeralCaps{18} and \textit{NuSTAR} not intersecting at the same point, nor agreeing with the spectral fit value. 

To determine the multi-thermal properties we calculate the emission measure distribution (EMD; the line-of-sight differential emission measure multiplied by the temperature bin width, in units of cm$^{-5}$) using the regularized inversion approach of \cite{hannah_multi-thermal_2013}. Both AIA and \textit{NuSTAR} data were used and the resulting EMD curves, and uncertainty regions, are shown in Figure~\ref{fig:emd}.

We find that, in Figure~\ref{fig:emd}, the calculated EMDs are consistent with the values obtained from the spectral fits and the loci curve upper boundaries. The EMD indicates a sharp edge at the quiescent AR/pre-flare spectral fitting values (Figure~\ref{fig:emd}, left panel) and a smoother drop in hotter material during the microflaring time (right panel). As the microflare heats the plasma an excess of material appears at temperatures where the ``tail'' of the pre-flare plasma falls off quickly. This behavior is similar to what has been found for significantly larger microflares, also observed in EUV and X-rays \citep{athiray_foxsi-2_2020}. The pre-flare time EMD in Figure~\ref{fig:emd} (left panel) again shows the importance of higher-energy X-ray spectroscopy when trying to robustly determine the presence of hot material in non-flaring ARs, as highlighted in previous studies \citep{reale_comparison_2009, schmelz_like_2009, ishikawa_detection_2017}. 

\begin{figure*}
\gridline{\fig{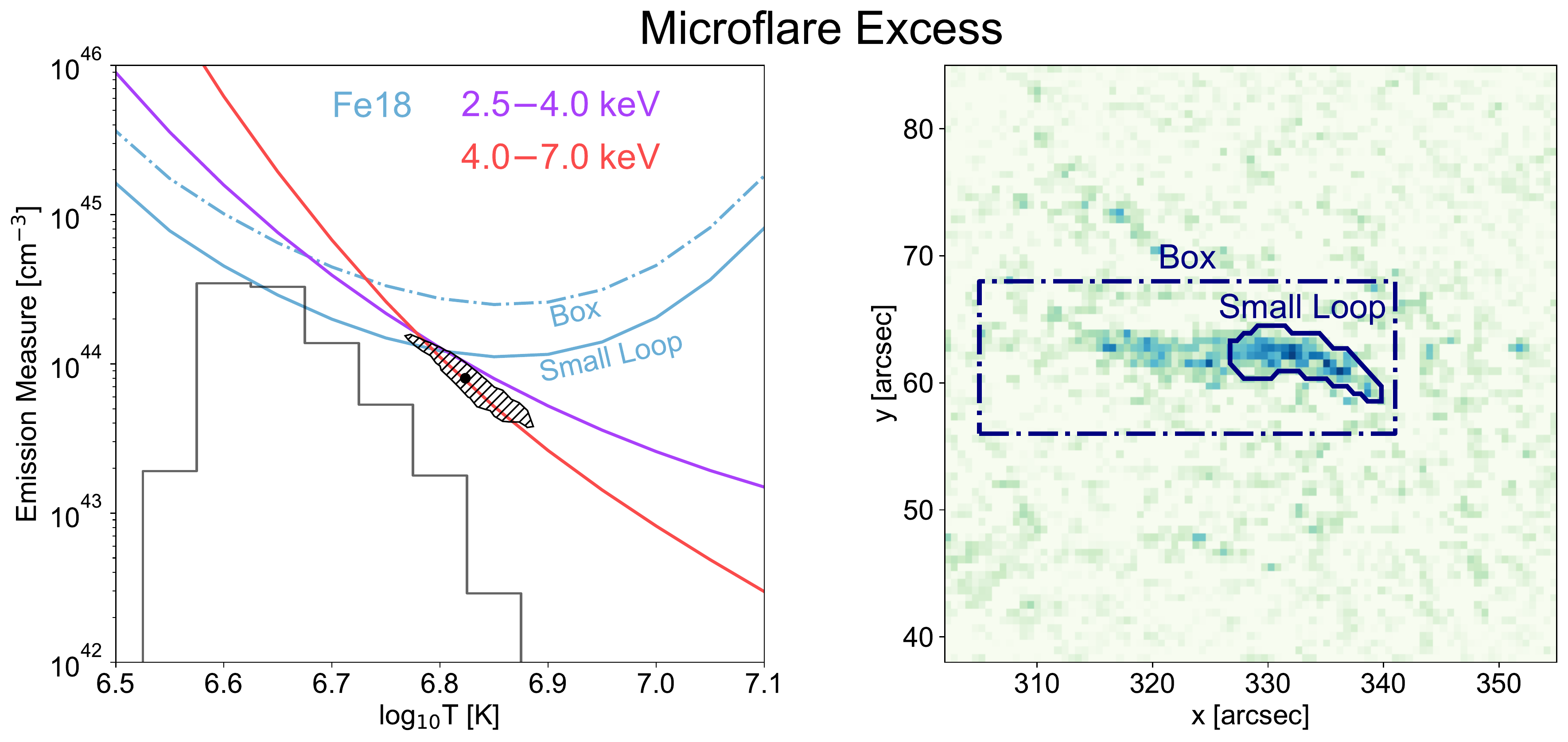}{0.89\textwidth}{}}
    \vspace{-0.8cm}
\caption{Microflare excess \textit{NuSTAR} (purple, red) and \textit{SDO}/AIA Fe~\RomanNumeralCaps{18} (blue) EM loci curves in comparison to the microflare excess EMD (gray). \textit{SDO}/AIA Fe~\RomanNumeralCaps{18} loci were obtained for two different source regions: ``Box" and ``Small Loop" indicated in the right panel. The spectral fit value for the microflare excess is shown in the loci plot with a black dot and 90\% confidence region where errors on all other quantities are omitted due to their large magnitude.
\label{fig:emdExcess}}
\end{figure*}

By subtracting the pre-flare emission from the microflare---isolating the microflare heated plasma---the loci curves show more consistent behavior with the spectral fit excess parameters (6.7~MK and 8.0$\times10^{43}$~cm$^{-3}$), again, indicating similar emission seen by both observatories (Figure~\ref{fig:emdExcess}, left panel).

Figure~\ref{fig:emdExcess} (left panel) displays further evidence that the microflare emission is from the right-hand side of the region assumed in Section \ref{sec:lc&im}. The Fe~\RomanNumeralCaps{18} (blue, solid) loci curve from the ``Small Loop" area is far more consistent with the \textit{NuSTAR} loci curves and spectral fit value compared to the larger ``Box" region loci curve (blue, dashed-dotted). This indicates that during the microflare time the excess emission is observed from this ``Small Loop" region and that the pre-flare emission was from a larger fraction of the AR. 

Further support for this is seen when we compare the observed \textit{SDO}/AIA Fe~\RomanNumeralCaps{18} flux from these regions to the synthetic AIA flux, calculated from the \textit{NuSTAR} thermal parameters folded through the AIA response. Using the AIA flux from the ``Box" we find that \textit{NuSTAR} appears to detect $\sim$62\% of the emission observed by \textit{SDO}/AIA Fe~\RomanNumeralCaps{18} (synthetic flux: 1.20$^{+0.11}_{-0.09}$~DN~s$^{-1}$~pixel$^{-1}$, observed flux: 1.95$\pm$0.06~DN~s$^{-1}$~pixel$^{-1}$) from the quiescent AR. However, only $\sim$30\% of the microflare excess is observed by \textit{NuSTAR} (0.07$^{+0.06}_{-0.04}$~DN~s$^{-1}$~pixel$^{-1}$) compared to Fe~\RomanNumeralCaps{18} (0.23$\pm$0.09~DN~s$^{-1}$~pixel$^{-1}$), whereas the synthetic flux obtained for the microflare excess in the ``Small Loop" region (1.14$^{+1.03}_{-0.57}$~DN~s$^{-1}$ pixel$^{-1}$) is $\sim$69\% of the observed flux (1.66$\pm$0.16~DN~s$^{-1}$~pixel$^{-1}$). The smaller region is more consistent for the microflare excess with the temperature calculated and the Fe~\RomanNumeralCaps{18} response. Thus, it is determined that the ``Small Loop" region shown in Figure~\ref{fig:emdExcess} is the true microflaring loop. A volume of 1.9$\times10^{25}$~cm$^{3}$ and energy of 1.1$^{+0.2}_{-0.2}\times$10$^{26}$~erg is then recalculated for this smaller loop, lowering the already small upper limit given to the instantaneous energy release of this microflare.

\section{Summary and Conclusions} \label{sec:conclusion}
Using \textit{NuSTAR}, in conjunction with \textit{SDO}/AIA, we have identified the smallest thermal energy X-ray microflare yet detected within an AR. A typical quiescent AR/pre-flare temperature and emission measure ($\sim$3~MK and $\sim$10$^{46}$~cm$^{-3}$ respectively) was obtained when fitting a thermal model to the spectrum with the microflare excess reaching temperatures up to 6.7~MK and an emission measure of 8.0$\times10^{43}$~cm$^{-3}$. This is hotter and has a lower emission measure than most of the previously studied \textit{NuSTAR} microflares \citep{hannah_first_2016, hannah_joint_2019, glesener_nustar_2017, wright_microflare_2017}.

The microflare is estimated to have a thermal energy release of 1.1$^{+0.2}_{-0.2}\times$10$^{26}$~erg. This is the not the most spatially compact microflare, but it is the smallest thermal energy release from an X-ray microflare observed in an AR. This thermal energy is comparable to the small brightenings seen in high-resolution EUV observations of magnetically braided loops \citep{cirtain_energy_2013}. This shows that with \textit{NuSTAR} we are starting to detect the X-ray emission from the myriad of small brightenings seen in EUV, and are approaching events closer to nanoflare than microflare energies.

No non-thermal emission was detected; however, some electron acceleration could have occurred throughout the evolution of the microflare. Support for this comes in the form of the higher energy time profile (\textit{NuSTAR} 4--7~keV) rising earlier than the lower energy profiles (\textit{NuSTAR} 2.5--4~keV and \textit{SDO}/AIA Fe~\RomanNumeralCaps{18}). We found non-thermal upper limits that were consistent with not producing detectable emission, yet still capable of matching the heating rate in this microflare.

This tiny microflare was very evident in the X-ray data but harder to find in the EUV emission, highlighting the need for sensitive X-ray telescopes to study flares. It may be easier, however, to find more events of this scale within ARs, using this one as an example. This would further the investigation into how the flare frequency distribution of smaller flares compare to that of larger ones \citep{crosby_frequency_1993, hudson_solar_1991, hannah_microflares_2011}. 

Throughout the six $\sim$1~hour \textit{NuSTAR} observations on 2018 September 9--10 there was a multitude of microflares from AR12721. The statistics of these events will be the subject of another paper, furthering our understanding of the range of small flares possible.

\vspace{0.2cm}
This Letter made use of data from the \textit{NuSTAR} mission, a project led by the California Institute of Technology, managed by the Jet Propulsion Laboratory, funded by the National Aeronautics and Space Administration. These observations were supported through the NuSTAR Guest Observer program (NASA grant 80NSSC18K1744). This research has made use of SunPy v1.0.6, an open-source and free community-developed solar data analysis Python package \citep{sunpy_community_sunpypython_2015}. This research also made use of HEASoft (a unified release of FTOOLS and XANADU software packages) and \textit{NuSTAR} Data Analysis Software (NuSTARDAS). This Letter made use of the SolarSoft IDL distribution (SSW) from the IDL Astronomy Library.

K.C. is supported by a Royal Society Research Fellows Enhancement Award and I.G.H is supported by a Royal Society University Fellowship.

\bibliography{ref}{}
\bibliographystyle{aasjournal}


\end{document}